\documentclass[prd,twocolumn,showpacs,amsmath,amssymb,epsf,epsfig]{revtex4-1}
\usepackage{graphicx}
\usepackage{epsfig}
\usepackage{amssymb}
\usepackage{graphicx}
\usepackage{dcolumn}
\usepackage{bm}
\newcommand{\ba}{\begin{array}}
\newcommand{\ea}{\end{array}}
\usepackage{graphicx}
\usepackage{dcolumn}
\usepackage{bm}
\def\br{\begin{eqnarray}}
\def\er{\end{eqnarray}}
\def\be{\begin{equation}}
\def\ee{\end{equation}}

\def\({\left(}
\def\){\right)}

\begin{document}
\title{Deciphering the minimum of energy of some walking technicolor models}

\author{A.~Doff$^1$ and A.~A.~Natale$^2$}
\affiliation{$^1$Universidade Tecnol\'ogica Federal do Paran\'a - UTFPR - COMAT
Via do Conhecimento Km 01, 85503-390, Pato Branco - PR, Brazil \\
$^2$Instituto de F\'{\i}sica Te\'orica, UNESP - Univ. Estadual Paulista,
Rua Dr. Bento T. Ferraz, 271, Bloco II,
01140-070, S\~ao Paulo - SP,
Brazil }

\date{\today}

\begin{abstract}
There are quasi-conformal theories, like the Minimal and Ultraminimal Technicolor models, which may break dynamically the gauge symmetry
of the Standard Model and at the same time are compatible with electroweak precision data. The main characteristic of this type of models is their fermionic content in one or more higher dimensional representations, therefore it is not immediate to know which model leads to the most attractive channel or the minimum vacuum energy state. We discuss the effective potential for composite operators for these models, verifying that their vacuum energy values are different, with the Ultraminimal model having a deeper minimum of energy. 
\end{abstract}


\maketitle

\par The nature of the Higgs boson is one of the most important problems in particle physics, and there are many questions that may be answered in the near future by the LHC experiments, such as: Is the Higgs boson, if it exists at all, elementary or composite, and what are
the symmetries behind the Higgs mechanism.  The possibility that the Higgs boson is a composite state instead of an elementary one is more akin to the phenomenon of spontaneous symmetry breaking  that originated from the effective Ginzburg-Landau Lagrangian, which can be derived 
from the microscopic BCS theory of superconductivity describing the electron-hole interaction (or the composite state in our case). This dynamical origin of the spontaneous symmetry breaking has been discussed with the use of many models, being the most popular one the technicolor (TC) model \cite{sannino}. 
  
  Unfortunately we do not know the dynamics that form the scalar bound state, which  should play the role of the Higgs boson in the standard model symmetry breaking. Most of the models for the spontaneous symmetry breaking of the standard model based on composite Higgs boson system depend on specific assumptions about the dynamics responsible for the bound state formation \cite{hs}, and the work in this area try to find the TC dynamics dealing with the particle content of the theory, in order to obtain a technifermion self-energy that does not lead to phenomenological problems as in the scheme known as walking technicolor \cite{walk}. These are theories where the incompatibility with the experimental data has been solved,
making the new strong interaction almost conformal and changing appreciably its dynamical behavior. We can obtain an almost conformal TC theory, when the fermions are in the fundamental representation, introducing a large number of TC fermions ($n_F$), leading to an almost zero $\beta$ function and flat asymptotic coupling constant. The cost of such procedure may be a large S parameter \cite{peskin} incompatible with the high precision electroweak
measurements. 

TC models with  fermions in other representations than the fundamental one, like happens in the Minimal \cite{sannino1} (MWT) and Ultraminimal \cite{sannino2} (UMT) TC models, are possible
viable models without conflict with the known value for the measured S parameter, which may be calculated assuming
valid the perturbative expressions for such parameter. These models have some phenomenological
differences \cite{sannino3}, although their fermionic content is not totally different and it is possible to have
even more extensions of this type of models \cite{sannino4}. They also have a different number of composite scalar
particles as well as different couplings among themselves \cite{we}. \linebreak
\hspace*{0.4cm}There is a striking difference between models based on fundamental or composite scalar bosons. In the
case of a fundamental scalar boson we just have a scalar potential with a mass and coupling constant 
conveniently adjustable to
provide the correct gauge symmetry breaking of the Standard Model. In the case of a composite scalar 
we do have a gauge theory at some energy scale with some fermionic content, and everything should be
calculable in terms of these quantities, although, due to the non-perturbative aspects of the symmetry
breaking, it is much more difficult to obtain precise evaluations of the physical parameters. 
In this work we will investigate another  characteristic of these models, which is the value of the state of minimum energy (the most attractive channel), i.e. discover which model leads to the tightest bound states. This
type of information can be obtained with the use of an effective potential for composite operators \cite{cornwall1},
and this is a more involved quantity to compute when the theory has fermions in several (and higher dimensional) representations, because it
is not just a matter of comparing Casimir operator eigenvalues as is usually performed for a gauge group with an unique fermionic representation. \\
\hspace*{0.4cm}In the sequence we introduce the Minimal and Ultraminimal TC models, discuss the fermionic
self-energies solutions for these models, and compute the vacuum energy with these solutions. We also use a standard walking (WT)
theory (by standard we mean a theory with fermions only in the fundamental representation) to compare the 
different minima of energy of these quasi-conformal theories.
The MWT model is based on a $SU(2)$ gauge group with two adjoint fermions \cite{sannino1}
\vspace*{-0.3cm}
\begin{equation}
Q^a_L  =	\left( \ba{c}
	      		U^a\\
	      		D^a
              \ea \right)_L  \;\; \; , \;\;\; U^a_R , \;\;\; D^a_R , \;\;\; a=1, 2, 3,
\label{eq10}
\end{equation}
where $a$ is the $SU(2)$ adjoint color index and the left-handed fields correspond to three ($SU(2)_L$) weak
doublets. The UMT model is based on a two colors group with two fundamental Dirac flavors 
$SU(2)_L\times U(1)_Y$ charged described by \cite{sannino2}
\begin{equation}
T_L  =	\left( \ba{c}
	      		U\\
	      		D
              \ea \right)_L  \;\; \; , \;\;\; U_R , \;\;\; D_R ,
\label{eq20}
\end{equation}
and also two adjoint Weyl fermions indicated by $\lambda^f$ with $f=1,2$, where these fermions are  singlets under  $SU(2)_L\times U(1)_Y$. 

The near conformal behavior for these models can be observed looking at the zero of the two-loop $\beta (g^2)$ function, which is given by
$\beta (g) = -\beta_0 \frac{g^3}{(4\pi)^2} - \beta_1 \frac{g^5}{(4\pi)^4}$,
where $ \beta_0 = (4\pi)^2 b= \frac{11}{3} C_2(G) - \frac{4}{3}T(R)n_F (R)$ and 
$\beta_1 = \left[\frac{34}{3}C_2^2(G)-\frac{20}{3}C_2(G)T(R)n_F -4C_2(R)T(R)n_F \right]$.
Where $C_2(R)I=T^a_RT^a_R$, $C_2(R)d(R)=T(R)d(G)$, $d(R)$ is the dimension of the representation $R$ and $G$ indicates the adjoint representation. It is interesting to compare the leading term of the $\beta$ function for the different models (indicated respectively by $b_{mi}$ and $b_{um}$, while the one of a simple walking TC theory is denoted by $b_w$). In the case of an $SU(2)$ gauge group with $8$ Dirac fermions we have $b_w = 2/16\pi^2$, while in the Minimal walking model we obtain the same coefficient with only $2$ fermions ($b_w=b_{mi}$)! The main difference among these models appears when we compute the S parameter whose
perturbative expression (in the massless limit) is
\be
S=\frac{1}{6\pi}\frac{n_F}{2}d(R) \,\, .
\label{s1}
\ee
The data requires the value of the $S$ parameter
to be less than about 0.3. According to the ``naive'' perturbative
estimate of Eq.(\ref{s1}) this requirement is indeed met for MWT (and also
for UMT). Early models, with fermions only in the fundamental
representation, needed a quite large $n_F$ to have a walking behavior, giving a
perturbative estimate of $S$ in contradiction with data. Sannino and collaborators have extensively advocated the advantage of working with higher dimensional fermionic representations. In the Refs.\cite{sannino1,sannino2,sannino3,MWT,UMT} walking  TC models are introduced with the advantage of a small number of technifermions and in conformity with high precision standard model data.
  
\par  In the Ref.\cite{dn1} we introduced a very general ansatz for the technifermion self-energy that interpolates between all  known forms of technifermionic self-energy. As we vary one parameter ($\alpha$) in our ansatz for the technifermionic self-energy, we go from the standard operator
product expansion (OPE) behavior of the self-energy to the one predicted by the extreme limit of a walking technicolor dynamics. The form of this ansatz  is reproduced  below  
\be 
\Sigma_{A}(p^2) \sim  \Lambda_{{}_{TC}}\left( \frac{ \Lambda^2_{{}_{TC}}}{p^2}\right)^{\alpha}\left[1 + a\ln\left(p^2/ \Lambda^2_{{}_{TC}} \right) \right]^{-\beta} .
\label{eq1}
\ee
\noindent  In this expression the standard OPE behavior for $\Sigma (p^2)$ is obtained when $\alpha \rightarrow 1$, whereas the extreme walking technicolor solution is obtained when $\alpha \rightarrow 0$.  We identify  $a \equiv bg^2_{{}_{TC}}$,  $\beta \equiv \gamma_{{}_{TC}}\cos(\alpha\pi) $ with $\gamma_{{}_{TC}} = \gamma = 3c/16\pi^2b$,  and  $c$ is the quadratic Casimir operator given by 
$
c = \frac{1}{2}\left[C_{2}(R_{1}) +  C_{2}(R_{1}) - C_{2}(R_{3})\right].
$
\noindent  $C_{2}(R_{i})$ are the Casimir operators for technifermions in the representations  $R_{1}$ and $R_{2}$ that condensate in the representation $R_{3}$, $b$ is the coefficient of the $g^3$ term in the technicolor $\beta (g)$ function.

\par  The  TC scale ($\Lambda_{{}_{TC}}$) is related to the technicolor condensate by $\langle \bar{\psi} \psi\rangle_{{}_{TC}} \approx \Lambda^{3}_{{}_{TC}}$ and  we can describe  the TC scale  in terms of measurable quantities and of group theoretical factors of the strong interaction responsible for forming the composite scalar boson. In the extreme limit of a walking technicolor dynamics we expect to have\cite{dn2}
\be 
\Lambda_{{}_{TC}} = v\left(\frac{8 \pi^2  a(2\gamma -1)}{N_{TC}n_F} \right)^{1/2}
\label{eq2}
\ee 
\noindent where $v \sim 246 GeV$ is the standard model VEV. 

\par The effective potential for composite operators is given by the following expression \cite{cornwall1}
\be
V(S,D) = - \imath \int \frac{d^4p}{(2\pi)^4}Tr ( \ln S_0^{-1}S - S_0^{-1}S + 1)
 +\,\,V_2(S,D),
\label{efpot}
\ee 
\noindent  where  $S$ and $D$ are the complete propagators of fermions and gauge bosons and $S_0$, $D_0$, are  the corresponding bare propagators.  The function $V_2(S,D)$ is the sum of two-particle irreducible vacuum diagrams, which, in the leading Hartree-Fock approximation, is depicted in the Figure (1).  $V_2(S,D)$ can be represented  analytically by 
\be
\imath V_2(S,D) = - \frac{1}{2} Tr(\Gamma S \Gamma S D) \,\, ,
\ee 
\begin{figure}[t]
\centering
\includegraphics[width=0.6\columnwidth]{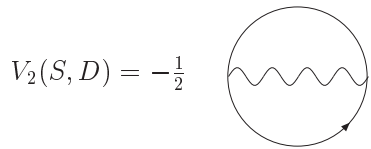}
\vspace*{-0.3cm}
\caption[dummy0]{Leading order contribution to  $V_2(S,D)$.}
\label{fig1}
\end{figure}
\noindent  where, for simplicity,  we have not written the gauge and Lorentz indices, as well as the momentum integrals and we represent the fermion proper vertex by $\Gamma$. 

\par We want to determine the vacuum expectation value obtained with the fermionic self-energy that is given by Eq.(\ref{eq1}), when $\alpha \approx 0$,  for the (MWT), (UMT)  and (WT) models. 
However, it is better to compute the vacuum energy density, which is given by the effective potential calculated at minimum subtracted by its perturbative part which does not contribute to dynamical mass generation\cite{cornwall1,castorina}
\be
\langle \Omega \rangle = V_{min}(S,D) - V_{min}(S_p,D_p),
\label{omega}
\ee 
\noindent where we indicate in the expression above the perturbative counterpart of $S$ and $D$ respectively by $S_p$, $D_p$.
 $V_{min}(S,D)$ is obtained substituting Eq.(\ref{eq1}) into Eq.(\ref{efpot}), assuming $\alpha \approx 0$. The complete fermion propagator $S$ is related to the free propagator by the equation $S^{-1} = S_0^{-1} - \Sigma$, with $S_0 = \imath /\not \! p$, and in the chiral limit $S_p = S_0$. 
We chose to work in the  Landau gauge for simplicity and after going to Euclidean space,  we find that $\Omega_{min} \equiv \langle \Omega \rangle$ 
and is equal to\cite{castorina}
\br
\hspace*{-0.5cm}\Omega_{min}  = -2N_{TC}n_F\!\!\int\!\!\! \frac{d^4p}{(2\pi)^4}\left[ \ln ( \frac{p^2 + \Sigma^2}{p^2} ) - \frac{\Sigma^2}{p^2 + \Sigma^2} \right]\!\!. 
\er
\par We can still expand  $\Omega_{min}$ in powers of $\Sigma^2/{p^2}$, so that
\be
\Omega_{min}  \approx  - N_{TC}n_F\int \frac{d^4p}{(2\pi)^4} \frac{\Sigma^4}{p^4}.
\label{eq7}
\ee
\par To obtain an analytical formula for the  vacuum energy density  we will  make the
substitution $x \rightarrow \frac{p^2}{\Lambda^2_{TC}}$ in the Eqs.(\ref{eq1}) and (\ref{eq7}),  and use the following   Mellin transform\cite{cs}
\be
\left[ 1 + \kappa \ln {x} \right]^{-\epsilon} =
\frac{1}{\Gamma ({\epsilon})}\int_0^\infty d\sigma \, e^{-\sigma}
\left( {x} \right)^{-\sigma \kappa} \sigma^{\epsilon - 1}
\label{mt}
\ee 
\noindent  that  will simplify considerably the calculation.  In this Mellin transform  we identified $\kappa =  a$ and $\epsilon = 4\beta$. Then, after we  substitute Eq.(\ref{eq1}) in to (\ref{eq7}), and perform  the integration  we obtain for $\alpha \approx 0$\cite{dn1}
\br
\Omega_{min}  = -\frac{\Lambda^4_{TC}}{16\pi^2 a}\frac{N_{TC}n_F}{(4\gamma - 1)}\left[1 - \frac{4\alpha}{a}\frac{1}{(4\gamma  - 2)} + O(\alpha^2)...\right].\nonumber \\ 
\label{pot}
\er 
\par In Table I we show the values  of the  coefficients $ a =  bg^2_{{}_{TC}}$ and  $\gamma_{i}$, for  technifermions in the fundamental representation $(i = F)$ or adjoint $(i = G)$,  obtained for  the MWT model and for a conventional WT model based on $SU(2)_{TC}$.

\begin{table}[t]
\begin{ruledtabular}
\begin{tabular}{cccccc}
 Model &  $a$ &  $\gamma_F$  &   $\gamma_G$ & $n_{F}(D)$ & $n_{F}(W)$ \\ \\
\hline
 WT  &  $0.21$ & $\frac{9}{8}$  & 0  &  8 &    0    \\ 
 MWT &  $0.08$ &  0  &  3 &   2  &  0   \\ 
 UMT &  $0.09$ &  $\frac{27}{40}$ &  $\frac{18}{10}$  &   2   &  2 \\ 
\end{tabular}
\end{ruledtabular}
\vspace*{-0.2cm}
\label{atab3}
\caption{ Values  for the  coefficients $a$ and  $\gamma$ obtained for  MWT, UMT and WT models. In this table we denoted by $n_{F}(D)$ and $n_{F}(W)$ respectively the number of  Dirac and  Weyl fermions. }
\end{table}
     
\par  In the case of the UMT model we  cannot apply straightforwardly  Eq.(\ref{eq2}), because in this case we have two scalar composite bosons, that appear as mixed states formed by fermions in the fundamental and adjoint representation. In this model only the lightest  composite boson, that is mostly formed by technifermions in the fundamental representation, is the one that couples to the particles of the Standard Model. The UMT gap equation has two contributions, one with a Casimir operator for fermions in the fundamental representation and another with a different Casimir  operator for fermions in the adjoint representation, while it is the same $\beta$ function that governs the running of the coupling in the two contributions.   

\begin{figure}[t]
\vspace*{-0.5cm}
\hspace*{-1.0cm}
\centering
\includegraphics[width=0.7\columnwidth]{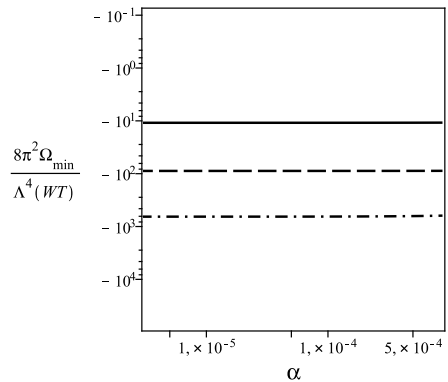}
\vspace*{-0.2cm}
\caption[dummy0]{ Behavior of  $\Phi(\Omega_{min}) \equiv \frac{8\pi^2\Omega_{min}}{\Lambda^4_{{}_{TC}(WT)}} $ for the WT (solid line), MWT (dashed) and UMT (dot-dashed) models  plotted as a function of the $\alpha$ parameter.
}
\label{WT0}
\end{figure}

It is opportune to remember that the gap equation lead to different chiral symmetry breaking scales when the fermions are in different representations, this has been observed, for instance, in QCD with quarks in the adjoint representation \cite{cornwall}, where the chiral transition may be slightly different from the confinement transition, which coincides with the chiral one for fermions in the fundamental representation.  We can expect that the masses and composite scalar wave functions will show a mixing but with scales quite close to the TC scale ($\Lambda_{TC}$), therefore  in the case of the UMT model we can suppose that  $\Lambda_{TC}(UMT) \approx \Lambda_{TC}(MWT)$; with this approximation (i.e. we assume the same
scales for the different models) we sum the contributions of the two different representations of the UMT model in order to compute Eq.(\ref{pot}).
 
It is also possible to relate  the TC scale  associated to the standard walking TC model ($\Lambda_{TC}(WT)$) with the scale of the minimal model considering Eq.(\ref{eq2}), which scale with the term between brackets
\be 
\frac{\Lambda_{{}_{TC}(WT)}}{\Lambda_{{}_{TC}(MWT)}} = \frac{\left(\frac{a(2\gamma -1)}{n_F} \right)^{1/2}_{WT}}{ \left(\frac{a(2\gamma -1)}{n_F} \right)^{1/2}_{MWT}} \,\, .
\ee 
\noindent Based on the above relation we can compute Eq.(\ref{pot}) for all models considering only a single scale( $\Lambda_{{}_{TC}(WT)}$), and
we can define for $SU(2)_{TC}$ the following quantity
\be 
\Phi(\Omega_{min}) \equiv \frac{8\pi^2\Omega_{min}}{\Lambda^4_{{}_{TC}(WT)}}.
\label{eqfi} 
\ee 
\par The value of Eq.(\ref{eqfi}) is plotted in Fig.(2) as a function of the $\alpha$ parameter. In this figure the solid line corresponds to $\Phi(\Omega_{min})$  obtained for the standard walking TC model, the  dashed line represents the corresponding result for the MWT model whereas the dot-dashed line is the result  obtained for the  UMT model. 

We have seen that the walking behavior can be obtained in many ways, for example  assuming a large number of technifermions in the fundamental representation or considering a small number of technifermions in higher dimensional representations. In this work  we consider  three different models that  lead to the walking behavior and  take the same form for the technifermion self-energy. However, analyzing Fig.(2) we verify that these three models have different values for the vacuum energy density. This result can be understood as follows,  as in the case of QCD with quarks in higher representations of $SU(3)_c$\cite{marciano},  technifermions in higher representations of $SU(2)_{{}_{}TC}$  naturally interact more strongly than conventional  technifermions and therefore lead to the deepest state of energy. These models can lead to a similar phenomenology, that in principle may be tested at the LHC, therefore, it is interesting to consider a criterion that  could be used to select which of these approaches may be the most promising to promote  the standard model symmetry breaking. 
Sannino and collaborators have extensively advocated the advantage of working with higher dimensional fermionic representations, in particular, in the refs.\cite{sannino1,sannino2,sannino3,MWT,UMT} are  introduced  the MWT and UMT models with a small number of technifermions and in conformity with high precision standard model data. 

\par In this work we proposed a mechanism to  select the most probable walking technicolor dynamics  assuming an energy criterion. We show that the Ultraminimal walking  TC models leads to a lower value for the minimum of the effective potential, or the formation of tightest bound states, with
the advantage of a small number of technifermions.

\vspace*{-0.2cm}
\acknowledgements
This research was partially supported by the Conselho Nacional de Desenvolvimento
Cient\'{\i}fico e Tecnol\'ogico (CNPq).

\end{document}